\documentclass{JAC2003}

\usepackage{graphicx}

\begin{document}

\title{The Quadratic Coefficient of the Electron Cloud Mapping}
\author{S. Petracca, A. Stabile, University of Sannio, Benevento (Italy),\\
T. Demma, Laboratoire de l'Accelérateur Linéaire
Dep. Accélélerateurs\\ Bât.209a
B.P.34 - 91898 ORSAY Cedex (France)}

\maketitle

\begin{abstract}
The \emph{Electron Cloud}  is an undesirable physical phenomenon which might produce single and multi-bunch
instability, tune shift, increase of pressure ultimately limiting the performance of particle accelerators.
We report our  results on the analytical study of the electron  dynamics.
\end{abstract}

\section{INTRODUCTION}
\label{sec:intro}
The electron cloud develops quickly as photons,
striking the vacuum chamber wall, knock out electrons  which are
subsequently accelerated by the beam and strike the chamber
again, producing further electrons in an avalanche process.
Most studies \cite{ecloud} performed so far were based on
 computer simulations (e.g. ECLOUD \cite{ecloud_sof}) taking into account
photoelectron production, secondary electron emission, electron
dynamics, and space charge effects, and providing a very detailed
description of the electron cloud evolution. In \cite{iriso1} it was shown that, for the typical parameters
of  the \emph{Relativistic Heavy Ion Collider} (RHIC), the evolution of
the longitudinal electron cloud density evolution  from bunch to bunch
can be  described (locally, i.e., at an arbitrary chosen point along
the beam path)  specific by  a simple "cubic map" of the form:

\begin{equation}
\rho_{m+1}\,=\,\alpha \, \rho_m +\beta \,{\rho_m}^2+\gamma \,{\rho_m}^3
\label{eq:cubic_map}
\end{equation}
where $\rho_m$ is the average electron cloud density after the $m$-th
passage of the bunch. A similar  map  was next suggested and found to be
reliable also for the \emph{Large Hadron Collider} (LHC) \cite{demma}. The coefficients $\alpha$, $\beta$, $\gamma$ are
extrapolated from simulations, and are functions of the beam
parameters and of the beam pipe features. The linear term  describes the linear growth and the coefficient $\alpha$
is larger than unity in the presence of electron cloud formation. The quadratic term describes the space charge effects, and is negative reflecting
the concavity to the curve $\rho_{m+1}$ vs $\rho_m$. The cubic term corresponds to
a variety of subtler effects, acting as perturbations to the above simple
scenario.

\begin{figure}[htb]
\centering
\includegraphics*[width=7cm]{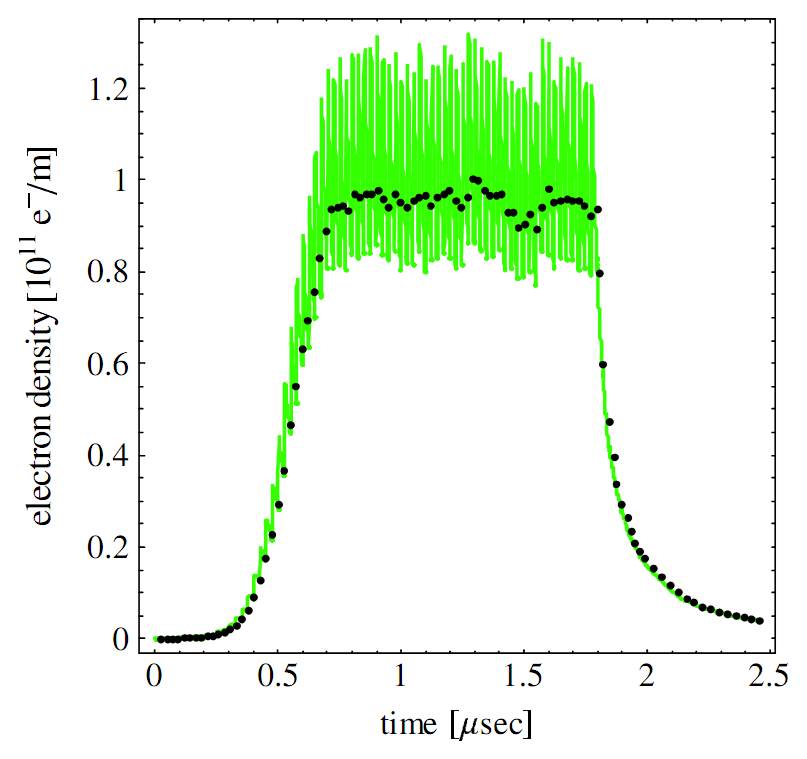}
\caption{Time evolution of the electron density (green line) computed with
ECLOUD. The black dots mark the average electron density between
two consecutive bunches obtained with the map formalism.
The machine/beam parameters used are
listed in Table 1.}
	\label{fig1}
\end{figure}

From Figure \ref{fig1} one can see that the bunch-to-bunch
evolution contains enough informations about the build-up or the
decay time, although the details of the line electron density
oscillation between two bunches are lost. The average longitudinal
electron density as function of time grows exponentially until the
space charge due to the electrons themselves produces a saturation
level. Once the saturation level is reached the average electron
density does not change significantly. The final decay corresponds
to the empty interval between successive bunches.

\begin{figure}[htb]
\centering
\includegraphics*[width=7cm]{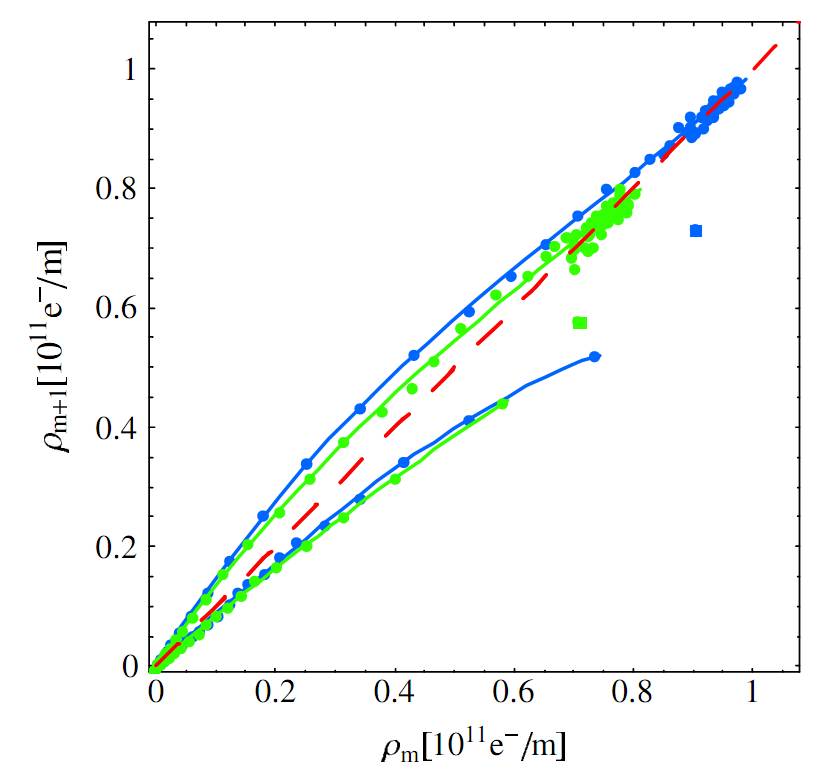}
\caption{Average longitudinal electron density  for different bunch populations
(green: $N_b\,=\,8*10^{10}$, blue: $N_b\,=\,16*10^{10}$). The lines
correspond to cubic fits.
The red line corresponds to the identity map
$\rho_{m+1}\,=\,\rho_m$. By comparison with Fig. 1, points above this line
describe the initial growth and saturation of the bunch-to-bunch evolution of the
electron density, those below describe the decay. The black line
represents the cubic fit of the points corresponding to the first
empty bunches.}
	\label{fig2}
\end{figure}

Fig. \ref{fig2} shows the behavior of the average electron
density $\rho_{m+1}$ as
function of the average electron density  $\rho_{m}$ for different
values of the bunch population (number of particles in a bunch, $N_b$).
\begin{table}[htb]
\centering
\begin{tabular}{||l|c|c|r||}
\hline\hline
Parameter& Quantity & Unit & Value \\
        \hline
        Beam pipe radius & $b$ & m & $.045$ \\
        Beam size & $a$ & m & $.002$ \\
        Bunch spacing & $s_b$ & m & $1.2$\\
        Bunch length & $h$ & m & $.013$\\
        Particles per bunch & $N_b$ & $10^{10}$& $4\,\div\,9$\\
\hline\hline
\end{tabular}
\caption{Input parameters for analytical estimate and ECLOUD
simulations.} \label{tab:tab1}
\end{table}
The markers in Fig. \ref{fig2} were obtained  from ECLOUD; the
lines are the cubic fits among these points.

The electron cloud dynamics can be roughly described
as follows: starting with a small
initial linear electron density, after some bunches the density
takes off and reaches the corresponding saturation line
($\rho_{m+1}\,=\,\rho_m$, red line) where the space charge effects, due
to the electrons in the cloud itself, take place. In this
situation, points corresponding to successive passages of filled
bunches, are in the same spot.

Until recently, the map coefficients have been extrapolated
from simulations, in a purely empirical way, so as to obtain
the best fit. An analytical expression of the linear coefficient $\alpha$ has been computed
in  a drift space \cite{iriso1},  and in the presence of a
magnetic dipole field \cite{demma}.

In this paper we summarize our recent results  \cite{demma2}
on the calculation of  an analytical expression for the quadratic coefficient
$\beta$ in  Eq. (1),  under the simple assumptions
of round chambers and free-field motion of the
electrons in the cloud. The coefficient $\beta$ turns out to depend
on few  beam and  machine parameters,  and
can be computed analytically once and for all,
saving a huge computational time compared to
 numerical simulations obtained using ECLOUD \cite{ecloud}.

This paper is accordingly organized as follows.
In  the next section we calculate the saturation  density of cloud electrons,
adopting a gaussian-like distribution for the secondary
electrons, producing an energy barrier near the
chamber wall. Later we deduce the formula
for the linear coefficient $\alpha$,  already given in \cite{iriso2},
and also for the  quadratic coefficient, which is a new result. Finally we report the conclusions.

\section{CLOUD SATURATION DENSITY}
\label{density}

Electrons in the cloud include both primary electrons,
generated by synchrotron radiation  at the pipe wall,
and secondary electrons,
produced  by  beam induced multi-pactoring. Primary electrons interact with the parent bunch,
and  are accelerated to a velocity
$v_p\!=\!2 c \bar{N}_b r_e/b$,
$r_e$ being the classical electron radius,
$b$ the pipe radius,
and $\bar{N}_b$ the linear particle density of
a longitudinally-uniform (coasting) beam having the same total
charge as the actual bunched beam,
\begin{equation}
\bar{N}_b\,=\,\frac{h}{h+s_b}N_b
\end{equation}
with  $s_b$   the intra-bunch spacing, and $h$ the  bunch length.

Secondary electrons
are produced with a low (typically a few $eV$)  energy  ${\mathcal{E}}_0$,
and move from the pipe wall,
with  velocity  $v_s\,=\,c \,\sqrt{2{\mathcal{E}}_0/mc^2}$
until the next bunch arrives. For large  $N_b$, $v_s < v_p$. Cloud buildup turns out to depend basically on two parameters \cite{Heifets}:
\vspace*{-1mm}
\begin{equation}
k=\frac{2\bar{N}_br_eh}{b^2} = b^{-1} v_p  \frac{h}{c},
\label{eq:k}
\end{equation}
and
\begin{equation}
\xi =\frac{h}{b}\sqrt{\frac{2{\mathcal{E}}_0}{mc^2}} = b^{-1} v_s \frac{h}{c},
\label{eq:xi}
\end{equation}
The parameters $k$ and $\xi$  are measures of the distances (in units of the pipe radius $b$)
traveled, respectively, by primary and secondary electrons during the bunch transit time. At low currents, $k\! \ll \!1$,  primary electrons interact with several bunches
before eventually  reaching the wall.
In the opposite extreme case,  $k\! >\! 2$, they travel from wall to wall
in a single bunch transit-time. The transition between the two regimes can be expected to occur at  $k\!\sim\! 1$.

For $k\!\!>\!\!1$  secondary electrons are confined in
a layer $\xi\!<\!r/b\!<\!1$  near the pipe wall,
and are wiped out of the region $0\!<\!r/b\!<\!\xi$
close to the beam by each passing bunch. Operating in the range of parameters
 ($k\!>\!1$ and $2-k\!<\!\xi\!<1$)
is thus clearly desirable to suppress the adverse effects
of the e-cloud on the beam dynamics \cite{Heifets}. In this range secondary electrons  create a  space-charge energy barrier
near the wall,  where they  are locked up,
and their density  grows until this barrier exceeds
their native energy  ${\mathcal{E}}_0$, viz.
\begin{equation}
-e\,V(1-\xi)\, \geq \,{\mathcal{E}}_0
\label{condition}
\end{equation}
where  $V$ is  the electric potential generated by the electron cloud,
and $-e$  is the electron charge.
The saturation condition corresponds to the equality in eq. (\ref{condition}).

To compute the potential in (\ref{condition}) we assume
a Gaussian radial dependence for the electron cloud charge density,
peaked at $r_0$, with std. deviation $\sigma$, viz.
\begin{equation}
\rho(r)\,=\,\rho_0\exp
\left[
-\frac{(r-r_0)^2}{2\sigma^2}
\right],
\label{radialdensity_1}
\end{equation}
where $\rho_0$ is fixed by the condition
\begin{equation}
2\pi h\int_a^b\rho(r)rdr\,=\,-N\,e
\end{equation}
$N$ being  the total number of electrons in the cylindrical shell
with radii $a,b$ and height $h$ around each bunch. Introducing the dimensionless quantities
$\tilde{a}=a/b$, $\tilde{r}=r/b$, $\tilde{r}_0=r_0/b$, $\tilde{\sigma}=\sigma/b$,
$g=\bar{N}_b/N$ and $V_0=Ne/2\pi\epsilon_0h$,
the (total) electric field and potential in the beam pipe can be written:
\begin{equation}
  \mathbf{E}(\tilde{r})\,=\,V_0
\left[g -\frac{F(\tilde{r})}{F(1)}
\right]\frac{\hat{\mathbf{r}}}{r},
\label{electricfield}
\end{equation}
\begin{equation}
V(\tilde{r})\,=\,
-V_0
\left[g\ln
\tilde{r}+\frac{G(\tilde{r})}{F(1)}
\right]
\label{radialpotential}
\end{equation}
where
\begin{equation}
F(\tilde{r})=\int_{\tilde{a}}^{\tilde{r}}
\exp\left[-\frac
{(y-\tilde{r}_0)^2}{2\tilde{\sigma}^2}
\right]y\,dy=
\end{equation}
and
\begin{equation}
G(\tilde{r})=\int_{\tilde{r}}^1\,\frac{F(y)}{y}dy,
\end{equation}

\begin{figure}[htb]
\centering\includegraphics[width=7cm]{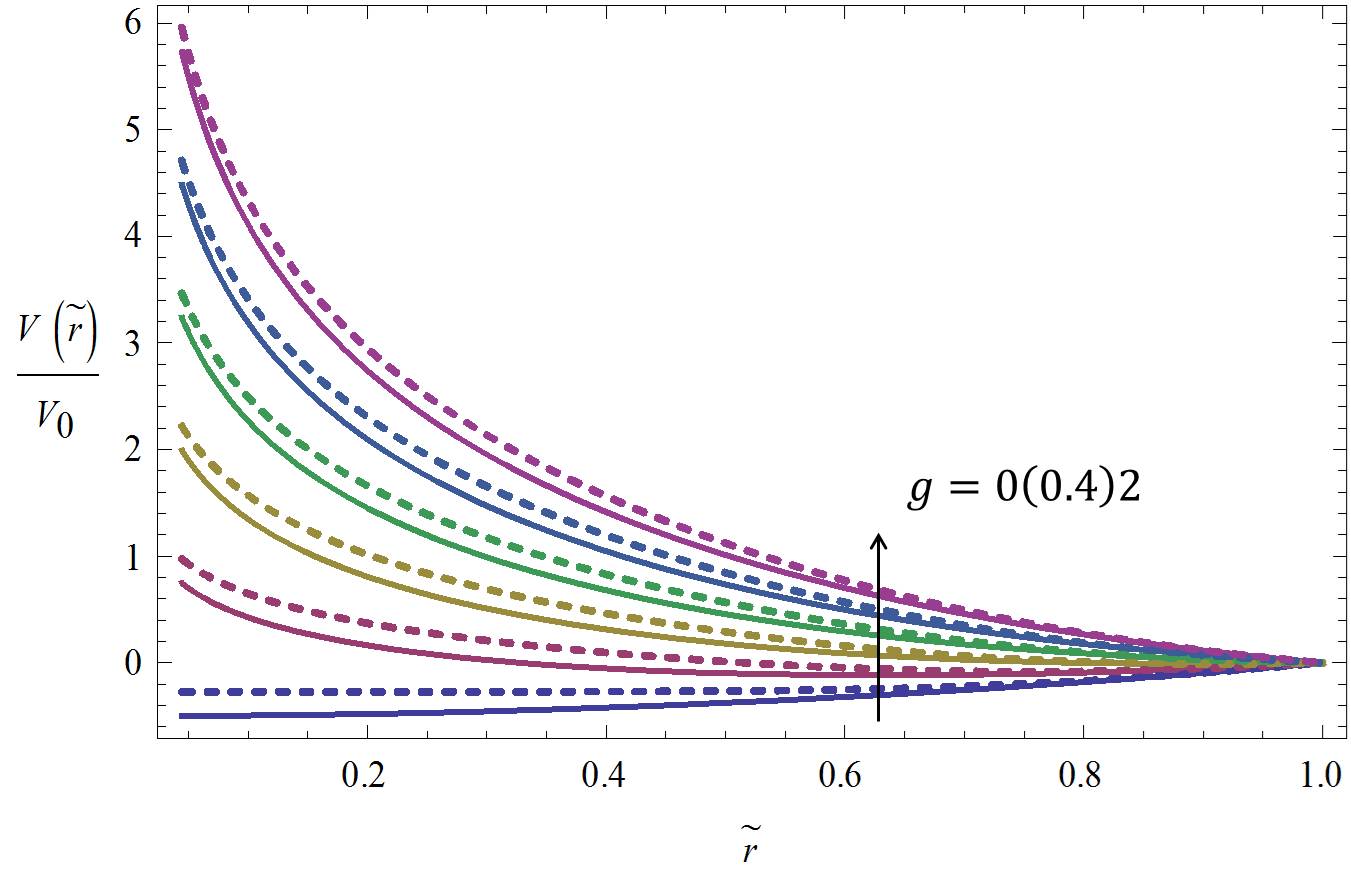}
\caption{The potentials in Eq. (\ref{radialpotential}) (dashed lines),
and (\ref{potentialuniform})  (solid lines)  as functions of $\tilde{r}$
for various values of $g$ and  $\tilde{a}\!=\!0.04$, $\tilde{r}_0\!=\!0.8$,
$\tilde{\sigma}\!=\!0.2$.}
\label{plotpotential}
\end{figure}

The limiting form of the potential for $\tilde{\sigma}\! \gg \!1$  (i.e., for a
uniform cloud charge density), and  $\tilde{a}\,\rightarrow\,0$ (vanishingly
thin beam) is:
\begin{equation}
V \rightarrow -V_0\biggl[g\ln \tilde{r}+\frac{1-\tilde{r}^2}{2}\biggr].
\label{potentialuniform}
\end{equation}
Figure \ref{plotpotential} displays the potential (\ref {radialpotential}) as
a function of $\tilde{r}$,  for various values of  $g$. The limiting form
(\ref{potentialuniform}) is also shown for comparison (dashed lines).
The potential  (\ref{potentialuniform})
is minimum at $\tilde{r}\,=\,\tilde{r}_m\,=\sqrt{g}$.
For $g\!>\!1$ it decreases monotonically with $\tilde{r}$
throughout the beam pipe $(0\!\leq\!\tilde{r}\!\leq\!1)$.
The condition $g\,=\,1$  corresponding  to  $N=N_b$,
i.e., to the well known condition of  neutrality \cite{Heifets}.
The potential (\ref{radialpotential}) obtained from the Gaussian
cloud density profile (\ref{radialdensity_1}) behaves similarly.

\begin{figure}[htb]
\centering\includegraphics[width=7.2cm]{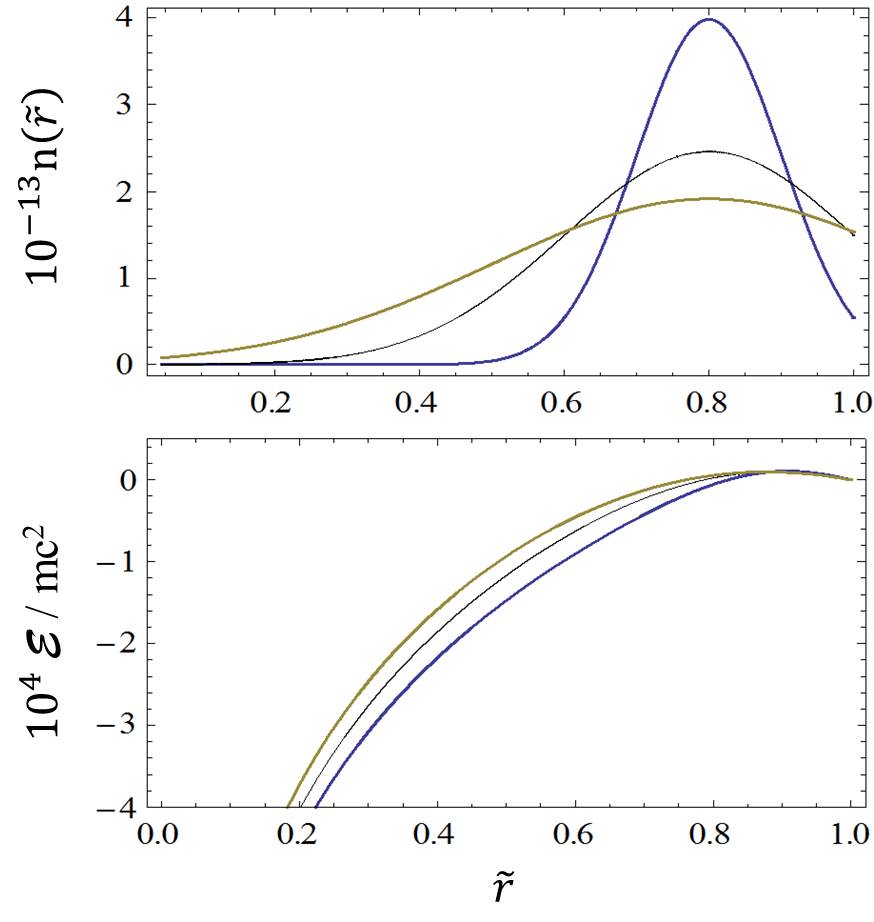}
\caption{The electron density  $n(\tilde{r})$  (solid line)
and  the energy barrier ${\mathcal E}(\tilde{r})$  (dashed line)
for $\tilde{\sigma}\,=\,0.1(0.1)0.3$, $\tilde{r}_0\,=\,0.8$, $N_b\,=\,6\cdot10^10$.}
\label{plot_density_energy}
\end{figure}

\begin{figure}[htb]
\centering\includegraphics[width=7cm]{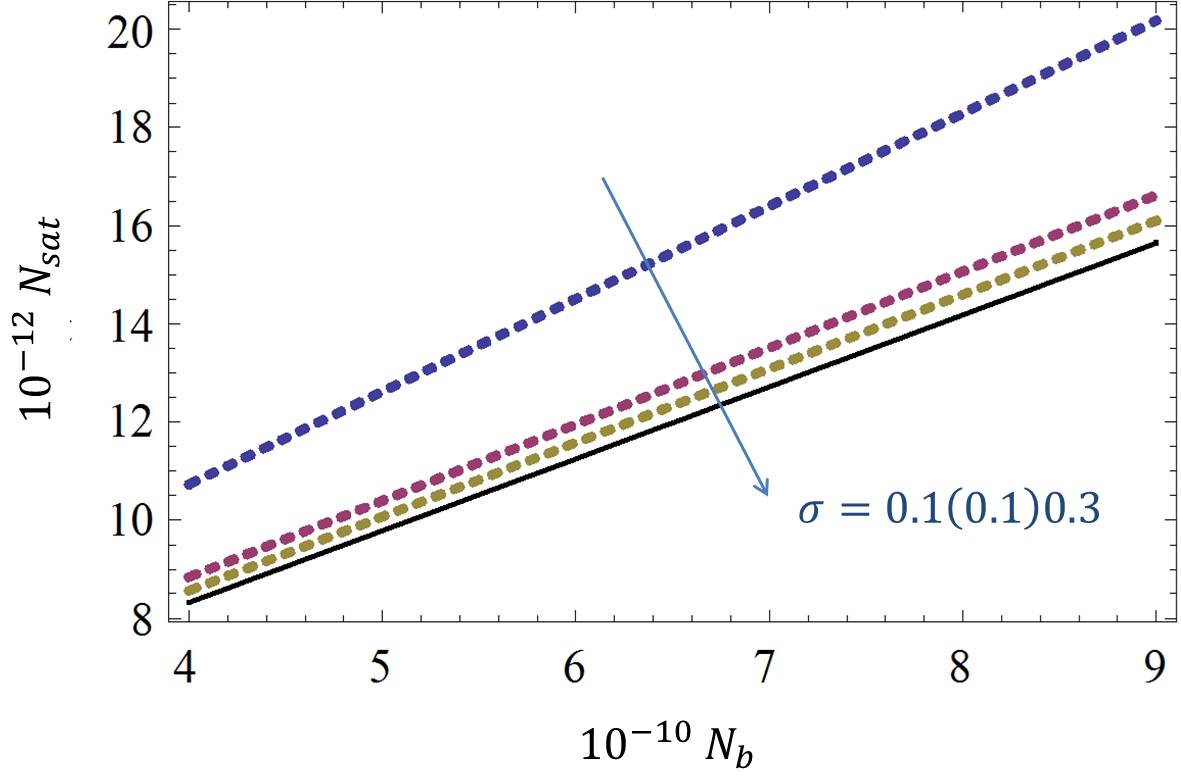}
\caption{The saturation electron density vs $N_b$.  Equation (\ref{nsatunif})
 (solid line)   and   eq.  (\ref{nsatgauss})  (dashed lines)
 for $\tilde{\sigma}\,=\,0.1(0.1)0.3$, with $\tilde{a}\,=\,0.04$, $\tilde{r}_0\,=\,0.8$. }
\label{plotdensitysaturation}
\end{figure}

The space-charge energy barrier
${\mathcal{E}}(r)\,=\,-e\,V(r)$
faced by the electrons originated at the walls
is  compared in  Figure  (\ref{plot_density_energy})
to the the electron density
$n(\tilde{r})=-\rho(\tilde{r})/e$.
%
It is seen that the position of the peak of the energy barrier
corresponds to the maximum concentration of the electrons,
and the barrier height goes to zero where the electron density vanishes. The saturation condition eq.(\ref{condition}) yields the following
critical number  of electrons in the cloud
\begin{equation}
N_{sat}\,=\, \frac{F(1)}{G(1-\xi)}
\left[
\frac{{\mathcal E}_0\,h}{2m\,c^2\,r_e}
-\bar{N}_b\ln(1-\xi)
\right]
\end{equation}
where $r_e$ is the classical radius of electron. Assuming the electrons as confined
in a cylindrical shell with inner radius $r_0-3\sigma$ and external radius $b$ the average saturation density can be written as
\begin{equation}
n_{sat}\,=\,\frac{\rho_{sat}}{-e}\,=\,
\frac{N_{sat}}{\pi h b^2 [
1-(\tilde{r}_0-3\tilde{\sigma})^2
]},
\label{nsatgauss}
\end{equation}
If the electrons are uniformly distributed
in the region $a\!\leq\!r\!\leq b$ we get
\begin{equation}
\bar{n }_{sat}\,=\,\frac{\bar{\rho}_{sat}}{-e}\,=\,\frac{\bar{N}_{sat}}{\pi h b^2 [1-\tilde{a}^2]}.
\label{nsatunif}
\end{equation}
In Figure \ref{plotdensitysaturation} we show the behavior of
the saturation densities (\ref{nsatgauss}) and (\ref{nsatunif}).

\section{ANALYTICAL DETERMINATION of COEFFICIENTS}
\label{coefficients}
\begin{figure}[htb]
\centering\includegraphics[width=7cm]{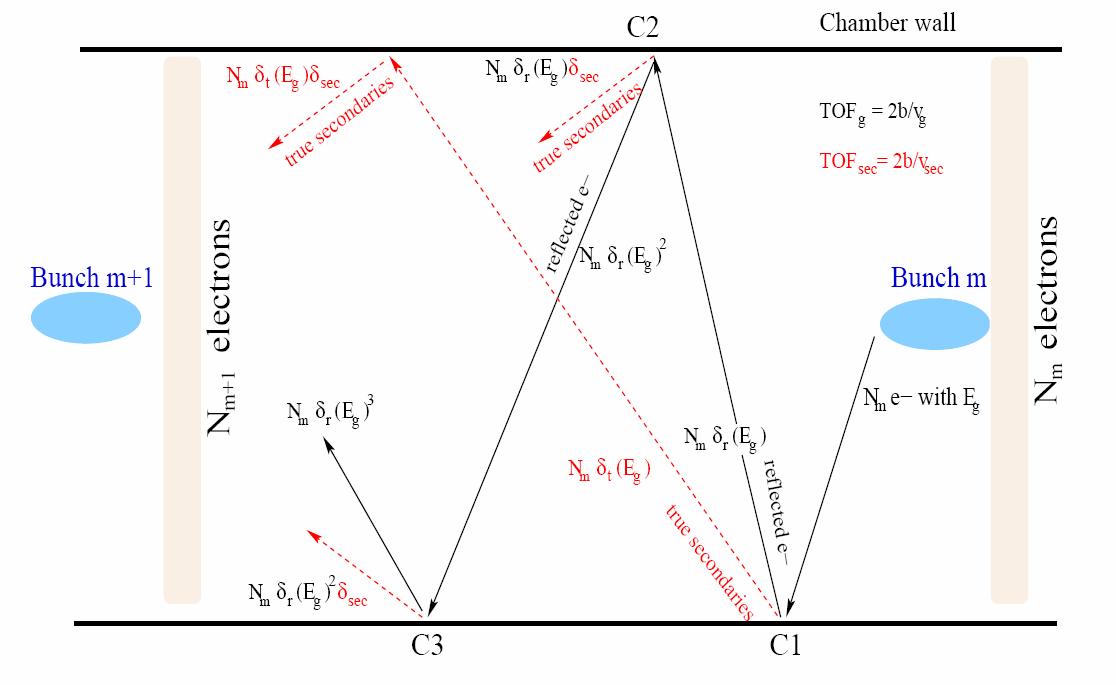}
\caption{
Sketch of the electron cloud evolution in the time interval between
two successive bunches. The $m$-th bunch accelerates the $N_m$
electrons  towards  the chamber wall with energy ${\mathcal{E}}_g$.
The first collision (C1) at the wall producwa two jets: the first one consisting
of $N_m\delta_r$  backscattered electrons (black line), with energy
${\mathcal{E}}_g$; the second one consisting of $N_m\delta_t$
secondary  electrons  (red dotted line) with low energy .
Before the $(m+1)$-th bunch arrives, these two jets undergo
further collisions, originating further jets.
The sum of over all these jets gives  $N_{m+1}$}
\label{fig:ecloud}
\end{figure}

Let  $N_m$ the total number of electrons in the cloud
at the passage of bunch-$m$. After the passage of the bunch
they are brought to an energy
\begin{equation}
{\mathcal{E}}_g\,=\,m_ec^2\frac{r_eN_b}{\sqrt{2\pi}h}
\left[
\log\biggl(\frac{b}{1.05a}\biggr)-\frac{1}{2}
\right]
\end{equation}
(see \cite{Berg} for derivation).
After a first collision with the pipe wall,
two electron jets are created:
a reflected (back-scattered) one, containing of $\delta_r({\mathcal E}_g)N_m$ electrons,
with  energy ${\mathcal{E}}_g$,
and a "true" secondary one, containing $\delta_t({\mathcal E}_g)N_m$ electrons,
with energy ${\mathcal E}_0$.
The quantities $\delta_{r,s}$  are referred to as $SEY$
(\emph{Secondary Emission Yield}) \cite{furman}),
and the process proceeds in cascade until the next bunch arrives, as sketched in Figure \ref{fig:ecloud}.
%
%
%
During the interval  $t_{sb}=h/\beta c$ preceding the passage of the next bunch,
electrons with  ${\mathcal E}\!=\!{\mathcal E}_g$  undergo a total number of collisions
with the pipe wall given by
\begin{equation}
S\,=\,\frac{t_{sb}-t_f({\mathcal{E}}_g)}{t_f({\mathcal{E}}_g)},
\label{eq:S}
\end{equation}
where  $t_f({\mathcal E_g})$  is the wall-to-wall flight time for an electron
with energy ${\mathcal E_g}$  (averaged over all possible angles w.r.t.
to the pipe axis), obtained from
\begin{equation}
t_f({\mathcal{E}})\,=\,\frac{4b}{\pi\sqrt{2{\mathcal{E}}/m_e}}
\end{equation}
Hence the total numer of high-energy electrons at the arrival of bunch-$(m\!+\!1)$ is
\begin{equation}
N_{m+1}({\mathcal E}_g)\!=\!N_m \delta^S_r({\mathcal E}_g).
\label{eq:reflected}
\end{equation}
The jet of low-energy secondary electrons originating after the $p-th$ collision
of the high energy electrons contains
\vspace*{-2mm}
$$
\delta_s({\mathcal E}_g)\delta_r^{p-1}({\mathcal E}_g)N_m
$$
electrons with energy ${\mathcal E}_0$. These low energy electrons, will undergo
a further number of collisions with the walls, before the next bunch arrives,
given by
\begin{equation}
k_p=\frac{t_{sb}-p\mbox{}t_f({\mathcal E}_g)}{t_f({\mathcal E}_0)}
\end{equation}
and at each collisions the number of these (slow) electrons
will change by a factor  $\delta_r({\mathcal E}_0)\!+\!\delta_r({\mathcal E}_0)$,
since both the reflected and secondary electrons will have
{\it the same} energy ${\mathcal E}_0$.
The total number of low-energy electrons at the
arrival of bunch-$(m+1)$ will accordingly be
\vspace*{-2mm}
$$
N_{m\!+\!1}({\mathcal E}_0)\!=\!N_m \delta_s({\mathcal E}_g)\cdot
$$
\vspace*{-5mm}
\begin{equation}
\cdot \sum_{p=1}^{S}
\delta^{p-1}_r({\mathcal E}_g)
\left[
\delta_t({\mathcal E}_0)\!+\!\delta_r({\mathcal E}_0)
\right]^{k_p}
\end{equation}
The number of (fast and slow) electrons $N_{m+1}$ on the arrival
of bunch-$(m+1)$  is  thus given by:
$$
N_{m+1}\,=\,N_{m}\left[{\delta_r}^S({\mathcal E}_g)\!+\!
\delta_t({\mathcal E}_g)\right.\cdot
$$
\vspace*{-4mm}
\begin{equation}
\cdot\left. \sum_{p=1}^S{\delta_r}^{p-1}({\mathcal E}_0){\delta_{tot}^{k_p}({\mathcal E}_0)}
\right]
\label{eq:dens}
\end{equation}
having set $\delta_{tot}=\delta_r+\delta_s$.
The above argument ignores saturation effects,
and can be consistently used to express the linear coefficient in the cubic map (\ref{eq:cubic_map})
in terms of the SEY coefficients, whose dependence from energy is known \cite{furman}, as follows
\vspace*{-2mm}
$$
\alpha\!=\!
\frac{N_{m+1}}{N_m}\!=\!
\delta_r^S({\mathcal E}_g) +
$$
\begin{equation}
+ \delta_t({\mathcal E}_g)
\delta_{tot}^{\eta}({\mathcal E}_0)
\frac{\delta_{tot}^{\eta S}({\mathcal E}_0)\!-\!\delta_r^{S}({\mathcal E}_0)}
{\delta_{tot}^{\eta}({\mathcal E}_0)\!-\!\delta_r({\mathcal E}_0)}
\end{equation}
where $\eta=t_f({\mathcal E}_g)/t_f({\mathcal E}_0)=({\mathcal E}_0/{\mathcal E}_g)^{1/2} \ll 1$.

\begin{figure}[htb]
\centering\includegraphics[width=7.4cm]
{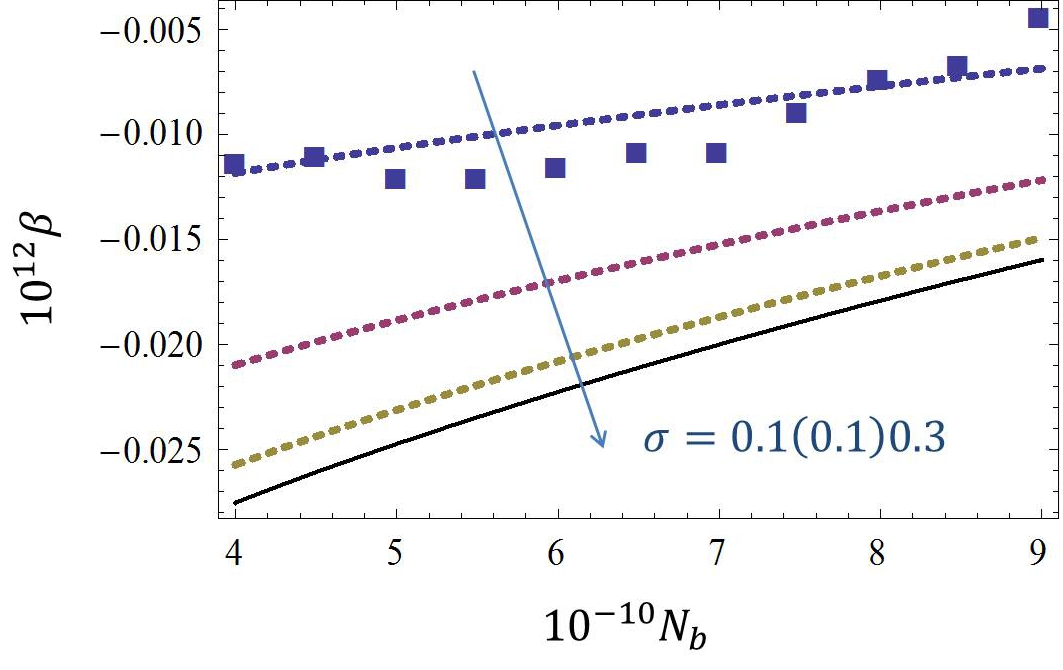}
\caption{The quadratic coefficient $\beta$  Eq. (\ref{quadratic_coeff})
versus $N_b$ (dashed lines), for
$\tilde{\sigma}=0.1(0.1)0.3$ and $\tilde{r}_0=0.95$,
and the corresponding values from ECLOUD  simulations (markers).
The solid line is obtained using in  (\ref{eq:beta}),
the uniform electron density (\ref{nsatunif}) .}
\label{plot_b}
\end{figure}

The coefficient $\beta$ in the cubic map (\ref{eq:cubic_map}) can now
be found by considering the saturation condition, where
\begin{equation}
\label{quadratic_coeff}
n_{sat}\,=\,\alpha\,n_{sat}+\beta\,{n_{sat}}^2\,\,\,\,
\end{equation}
yielding
\begin{equation}
\beta\,=\,\frac{1-\alpha}
{n_{sat}}
\label{eq:beta}
\end{equation}
Note that as $ {\mathcal E}_g \rightarrow \infty $, the quantity $S$
diverges, according to (\ref{eq:S}),
and hence the contribution of the high-energy electrons to eq. (\ref{eq:dens}) becomes negligible,
given that $\delta_r(\infty) < 1$  \cite{furman}.
We may accordingly use the saturation density of the secondary electrons,
derived in Section \ref{density},  to evaluate the quadratic coefficient
via  (\ref{eq:beta}).
%
%
%
%
In Figure  (\ref{plot_b}) we compare our analytic result for $\beta$
to the outcomes of simulations (ECLOUD code)
using the parameters in Table
\ref{tab:tab1}. A good qualitative agreement is obtained,
for the assumed Gaussian charge distribution.
%

\section{CONCLUSIONS}
\label{sec:conclu}
%
The main results in this paper can be summarized as follows.
A simple analytic form for the quadratic map coefficient $\beta$ has been derived,
and found to be in good  agreement compared to results obtained from
ECLOUD simulations. The map formalism can thus be easily applied
to determine safe regions in parameter space where the accelerator
can be operated without suffering from problems originated
by the electron clouds.


\begin{thebibliography}{99}
\itemsep=0ex


\bibitem{ecloud}  F. Zimmermann, "A Simulation Study of Electron-Cloud Instability and Beam-Induced Multipacting in the LHC,"  CERN  LHC-Project-Rept.  95  (1997).
\bibitem{ecloud_sof} http://ab-abp-rlc.web.cern.ch/ab-abp-rlc-ecloud/.
\bibitem{iriso1} U. Iriso, S. Peggs,  "Maps for Electron Clouds," Phys. Rev. ST-AB {\bf 8} (2005) 024403.
\bibitem{iriso2} U. Iriso, S. Pegg., "An Analytic Calculation of the Electron Cloud Linear Map Coefficient," Proc. EPAC '06, Edinburgh (UK), June 26-30 2006, paper MOPCH133.
\bibitem{demma} Th. Demma et al.,  "Maps for Electron Cloud Density in Large Hadron Collider Dipoles," Phys. Rev. ST-AB {\bf 10} (2007) 114401.
\bibitem{demma2}  Th. Demma et al., "E-Cloud Map Formalism: an Analytical Expression for Quadratic Coefficient,"
Proc. IPAC'10, Kyoto (JP), May 23-28 2010, paper TUPD037.
\bibitem{Heifets} S. Heifets, "Electron Cloud at High Beam Currents," SLAC-PUB-9584 (2002).
\bibitem{Berg} S. Berg, "Energy Gain in an Electron Cloud following the Passage of a Bunch," CERN LHC Proj. Note 97 (1997).
\bibitem{furman} M. Furman, M. Pivi, "Probabilistic Model for the Simulation of Secondary Electron Emission," Phys. Rev. ST-AB  {\bf 5} (2002) 12404.

\end{thebibliography}
\end{document}